\documentclass[12pt]{article}

\usepackage{epsfig}

\usepackage{amsmath} 
\usepackage{amssymb}
\usepackage{euscript}

\usepackage{cite}

\usepackage{alltt}

\newcommand{\Pcal}{{\cal P}}
\newcommand{\Peu}{\EuScript{P}}
\def\qi{{q_{i}}}
\def\qj{{q_{j}}}
\def\qibar{{\overline{q}_{i}}}
\def\qjbar{{\overline{q}_{j}}}
\def\qbar{{\overline{q}}}

\def\alfapi{\frac{\alpha_s}{\pi}}
\def\alfapi_t{\frac{\alpha_s}{2\pi}}

\begin{document}                     

\begin{titlepage}                     
\begin{flushright}
{\bf IFJPAN-IV-2007-7}
\end{flushright}

\vspace{1mm}
\begin{center}
{\LARGE\bf%
Solving the QCD NLO evolution equations \vspace{4mm}\\ 
with a Markovian Monte Carlo$^{\star}$%
}
\end{center}
\vspace{5mm}

\begin{center}
{\large\bf W.\ P\l{}aczek$^a$, K.\ Golec-Biernat$^{b,c}$, S.\ Jadach$^b$ }
{\rm and}
{\large\bf M.\ Skrzypek$^b$} 

\vspace{4mm}
{\em $^a$Marian Smoluchowski Institute of Physics, Jagiellonian University,\\
   ul.\ Reymonta 4, 30-059 Cracow, Poland.}\\ \vspace{2mm}
{\em $^b$Institute of Nuclear Physics, Polish Academy of Sciences,\\
  ul.\ Radzikowskiego 152, 31-342 Cracow, Poland.}\\ \vspace{2mm}
{\em $^c$Institute of Physics, University of Rzeszow,\\
   ul.\ Rejtana 16A, 35-959 Rzeszow, Poland.}
\end{center}
\vspace{15mm}

\begin{abstract}
We discuss precision Monte Carlo (MC) calculations for solving 
the QCD evolution equations up to the next-to-leading-order (NLO) level. 
They employ forward Markovian Monte Carlo algorithms,
which provide rigorous solutions of the above equations.
These algorithms are implemented in the form of the Monte Carlo program 
{\tt EvolFMC}.
This program has been cross-checked 
with independent, non-MC, programs ({\tt QCDNum16} and {\tt APCheb33})
and the numerical agreement at the level of $0.1\%$ has been found.
\end{abstract}

\vspace{5mm}
\begin{center}
{\em
Presented by W.\ P\l{}aczek at the Cracow Epiphany Conference\\
on Precision Physics and Monte Carlos for LHC,\\
4--6 January 2007, Cracow, Poland;\\
to be published in Acta Physica Polonica B. 
}
\end{center}

\vspace{5mm}
\begin{flushleft}
{\bf IFJPAN-IV-2007-7}\\
April~2007
\end{flushleft}

\vspace{5mm}
\footnoterule
\noindent
{\footnotesize
$^{\star}$%
 The project supported by EU grant MTKD-CT-2004-510126,
 realized in the partnership with CERN PH/TH Division and by the Polish
 Ministry of Scientific Research and Information Technology grant 
 No 620/E-77/6.PR UE/DIE 188/2005-2008.
}

\end{titlepage}                     

\section{Introduction}
\label{sec:intro}

Evolution equations of the quark and gluon distributions in a hadron,
known as the DGLAP equations,
derived in QED and QCD using the renormalization group or diagrammatic
techniques~\cite{DGLAP} can be interpreted probabilistically as a Markovian 
process, see e.g.\ Ref.~\cite{stirling-book}.
Such a process can be modeled using Monte Carlo methods.
The corresponding MC algorithm provides, in principle, an exact solution 
of the evolution equations for parton distribution functions (PDFs). 
In practice, the main limitation of such a solution is the size
of a generated MC sample, i.e. corresponding statistical errors of numerical
results. This is probably the main reason why this possibility
has not been exploited until recently.
Instead, alternative numerical methods and
programs solving the QCD evolution equations much faster than the Markovian
MC have been used, see e.g.~\cite{qcdnum16,APCheb33,blumleinn96}.

The feasibility of solving efficiently the DGLAP equations~\cite{DGLAP}
at the leading-order (LO) approximation with the Markovian MC was demonstrated 
for the first time in Ref.~\cite{Jadach:2003bu}.
The main conclusion of the above work was that the currently
available computer CPU power allows to solve efficiently and precisely
(at the per-mill level) the QCD evolution equations with the use of the
Markovian MC algorithm.
Of course, this method will always be slower in 
CPU time than non-MC techniques.
However, it has several advantages, such as: 
no biases and/or numerical instabilities related to finite grids of points, 
use of quadratures, decomposition into finite series of polynomials,
accumulation of rounding errors, etc. It is also more flexible in 
treatment of the PDFs (e.g. no need to split them into singlet and non-singlet 
components) and easier to extend to higher orders, new contributions, etc.
The above Markovian algorithm can form a basis of a final-state
radiation (FSR) parton shower MC program, which not only solves numerically
the evolution equations but also generates events in terms of parton
flavours and four-momenta.
Moreover, this algorithm is a starting 
point and a testing tool for various kinds of constrained MC 
algorithms being developed for the initial-state radiation (ISR),
see e.g.\ 
Refs.~\cite{raport04-06,Jadach:2005bf,Jadach:2005yq,Jadach:2007singleCMC}.

Here we briefly discuss the Markovian MC solution of the DGLAP 
evolution equations up to the next-to-leading order
in the perturbative QCD;
more details can be found in Ref.~\cite{Golec-Biernat:2006xw}.
The paper is organized as follows. In Section~2 we present a general structure
of the DGLAP equations and discuss their basic features up to the
next-to-next-to-leading order (NNLO). 
In Section~3 we briefly present the Markovian MC algorithm for parton-momentum
distributions. 
Numerical results from {\tt EvolFMC} at the NLO are presented 
in Section~4. They are compared with the results of non-MC program
{\tt APCheb33}. Comparisons with another non-MC program, {\tt QCDNum16},
are also briefly discussed. 
Finally, Section~5 contains the summary and outlook. 

\section{QCD evolution equations}
\label{sec:QCD-evolution}

The general form of the DGLAP evolution equations reads
\begin{equation}
\begin{split}
\label{eq:dglap1}
\frac{\partial}{\partial\ln\mu^2}\,\qi\:\:&=\:\:\sum_j\left(
P_{\qi\qj}\otimes \qj\, +\, P_{\qi\qjbar}\otimes \qjbar\right) 
+\, P_{\qi G}\otimes G\,,
\\
\frac{\partial}{\partial\ln\mu^2}\,\qibar\:\:&=\:\:\sum_j\left(
P_{\qibar\qj}\otimes \qj\, +\, P_{\qibar\qjbar}\otimes \qjbar\right) 
+\, P_{\qibar G}\otimes G\,,
\\
\frac{\partial}{\partial\ln\mu^2}\, G\:\:&=\:\:\sum_j\left(
P_{G\qj}\otimes \qj\, +\, P_{G\qjbar}\otimes \qjbar\right)
+\, P_{G G}\otimes G \,,
\end{split}
\end{equation}
where
$ \{q_1,\ldots , q_{n_f},~ \qbar_1,\ldots ,\qbar_{n_f},~ G\}(\mu,x)$
~--  quark, antiquark and gluon distributions;
$x$ -- Bjorken variable; $\mu$ -- hard scale,
(e.g. $\mu=\sqrt{Q^2}$ in DIS).

The integral convolution denoted by $\otimes$ involves only 
{\it longitudinal} momentum fractions:
\begin{equation}
\begin{split}
(P\otimes q)(\mu,x)= &
\int\limits_0^1dy \int\limits_0^1dz\,\delta(x-zy)\,P(\alpha_s,z)\, q(\mu,y) 
\\
= &
\int\limits_x^1\frac{dz}{z} \,P(\alpha_s,z)\,q\left(\mu,\frac{x}{z}\right)
\,.
\end{split}
\end{equation}

The {\em splitting functions} $ P(\alpha_s,z)$
depend on $\mu$ through the coupling  
constant $\alpha_s=\alpha_s(\mu)$:
\begin{equation}
\label{eq:splfun1}
P(\alpha_s,z)\,=\,
\underbrace{\frac{\alpha_s}{2\pi}\,P^{(0)}(z)}
_{\textrm{LO}}\,+\,
\underbrace{\left(\frac{\alpha_s}{2\pi}\right)^2 P^{(1)}(z)}
_{\textrm{NLO}}\,+\,
\underbrace{\left(\frac{\alpha_s}{2\pi}\right)^3 P^{(2)}(z)}
_{\textrm{NNLO}}\,+\ldots\,.
\end{equation}

From the charge conjugation and the $SU(n_f)$ symmetry 
the splitting functions $P$ have the following general structure
\begin{equation}
\begin{split}
\label{eq:splfun2}
P_{\qi\qj}\,=&\,P_{\qibar\qjbar}\,=\,\delta_{ij}P_{qq}^V\,+\,P_{qq}^S\,,
\\
P_{\qi\qjbar}\,=&\,P_{\qibar\qj}\,=\,\delta_{ij}P_{q\qbar}^V\,
+\,P_{q\qbar}^S\,,
\\
P_{\qi G}\,=&\,P_{\qibar G}\,=\,P_{FG}\,,
\\
P_{G\qi}\,=&\,P_{G\qibar}\,=\,P_{GF}\,.
\end{split}
\end{equation}

This leads to the basic form of the DGLAP evolution equations
\begin{equation}
\begin{split}
\label{eq:dglap2}
\frac{\partial}{\partial\ln\mu^2}\, \qi\,=&\,
P_{qq}^V\otimes\,\qi\,+\,P_{q\qbar}^V\otimes\,\qibar
\,+\,P_{qq}^S\otimes\sum_j\qj\,+\,P_{q\qbar}^S\otimes\sum_j\qjbar
\\
\, &+\, P_{F G}\otimes G\,,
\\
\frac{\partial}{\partial\ln\mu^2}\, \qibar\,=&\,
P_{q\qbar}^V\otimes\,\qi\,+\,P_{qq}^V\otimes\,\qibar\,+\,
P_{q\qbar}^S\otimes\sum_j\qj\,+\,P_{qq}^S\otimes\sum_j\qjbar
\\
\, &+\, P_{F G}\otimes G\,,
\\
\frac{\partial}{\partial\ln\mu^2}\, G\,=&\,P_{GF}\otimes 
\sum_j(\qj+\qjbar)
\\
\, &+\, P_{G G}\otimes G\,.
\end{split}
\end{equation}
Within a given  approximation  some splitting functions may
vanish  or be equal, e.g.\ at the LO:
$P^{V(0)}_{q\qbar}\,=\,P^{S(0)}_{q\qbar}\,=\,P^{S(0)}_{qq}\,=\,0\,,$
and at NLO:
$P_{qq}^{S(1)}\,=\,P_{q\qbar}^{S(1)}\,.$

\subsection{Singlet case}
The singlet PDF is defined as
\begin{equation}
\Sigma(\mu,x)\,=\,\sum_{j=1}^{n_f}\left[q_j(\mu,x)+\qbar_j(\mu,x) \right]\,.
\end{equation}
Introducing the notation
\begin{equation}
P_{FF}\!=\!P_{+}^{V}+n_f P_{+}^{S}\,,
\hspace{15mm}
P^{V,S}_{\,+}\!=\!P^{V,S}_{qq}\,+\, P^{V,S}_{q\qbar}\,,
\end{equation}
we obtain the following evolution equations for the quark-singlet and gluon 
distributions
\begin{equation}
\begin{split}
\frac{\partial}{\partial\ln\mu^2}\, \Sigma\,=&\,
P_{FF}\otimes\, \Sigma\, +\,(2n_f P_{F G})\otimes G\,,
\\
\frac{\partial}{\partial\ln\mu^2}\, G\,=&\,P_{GF}\otimes \Sigma\, 
+\, P_{GG}\otimes G\,.
\end{split}
\end{equation}
 The above splitting functions obey the general relations
\begin{equation}
\begin{split}
\int\limits_0^1 dz\,\{zP_{FF}(\mu,z) & +  zP_{GF} (\mu,z)\}\,
\\
& =
\int\limits_0^1 dz\,\{2n_f zP_{F G}(\mu,z)+zP_{GG}(\mu,z)\}\, 
=\,0\,.
\end{split}
\end{equation}
This leads to  the {\em momentum sum rule}
\begin{equation}
\int\limits_0^1dx \left\{x\Sigma(\mu,x)\,+\,xG(\mu,x)\right\}\,
=\,{\rm const}\,, 
\end{equation}
where ${\rm const} =1$ in the parton model.

\subsection {Non-singlet case}

The basic non-singlet PDF reads
\begin{equation}
V(\mu,x)\,=\,\sum_{j=1}^{n_f}\, \left[q_j(\mu,x)-\qbar_j(\mu,x)\right]\,,
\end{equation}
and its evolution equations is given by
\begin{equation}
\frac{\partial}{\partial\ln\mu^2}\,V\,=\,P^V_{NS}\otimes V\,,
\end{equation}
where the new splitting function
\begin{equation}
P^V_{NS} \!=\! P^{V}_{\,-}+n_fP^S_{\,-}\,,
\hspace{10mm}
P^{V,S}_{\,-}\!=\!  P^{V,S}_{qq}-P^{V,S}_{q\qbar}\,.
\end{equation}

The set of the splitting functions (the QCD kernels) usually represented 
in the literature reads
\begin{equation}
\{P^{V}_{\pm},\,P^S_{\pm},\,P_{FG},\,P_{GF},\,P_{GG}\}.
\end{equation}
$P^S_{+} = 0$ {at the LO,}
$P^S_{-} = 0$ {at the LO and at the NLO,}
{others} $\neq 0$ {at any order.}
Having the above splitting function one can write and solve the evolution 
equations  
in any of the presented forms.
In our Monte Carlo approach we work directly in the flavour space.
The general parton--parton transition matrix for a gluon and three quark 
flavours $(d,\,u,\,s)$ as well as its LO and NLO contributions are
given explicilty in Ref.~\cite{Golec-Biernat:2006xw}.

\subsection {Behaviour at $z\rightarrow 1$}

The splitting functions
$\{P^V_{\pm},\,P^S_{\,-},\,P_{GG}\} $
have  the following form
\begin{equation}
P(\alpha_s,z)\,=\,\frac{A(\alpha_s)}{(1-z)_+}\,+\,
B(\alpha_s)\,\delta(1-z)\,+\,\overline{P}(\alpha_s,z)\,.
\end{equation}
The functions $A(\alpha_s),\,B(\alpha_s)$ and 
$\overline{P}(\alpha_s,z)$ are calculated in powers 
of $\alpha_s$, e.g.
\begin{equation}
\overline{P}(\alpha_s,z)\,=\,\sum_{k=0}\alpha_s^{k+1}\,{D}^{(k)}(z)\,,
\end{equation}
where at the NLO and the NNLO the coefficients 
${D}^{(k)}(z)$ are logarithmically divergent:
\begin{equation}
{D}^{(k)}(z)\,=\,D_k\,\ln(1-z)\,+\,{\cal O}(1)\,.
\end{equation}

Similarly, the splitting functions $\{P_{FG},\,P_{GF}\}$
contain logarithmically divergent terms:
\begin{equation}
P(\alpha_s,z)\,=\,
\left\{
\begin{array}{ll}
{\cal{O}}(\alpha_s)~~~~~                          &  {\rm at~LO~(k=0)}\\
{\cal{O}}(\alpha_s^2\ln^2(1-z))~~~~~              &  {\rm at~NLO~(k=1)}\\
{\cal{O}}(\alpha_s^3\ln^4(1-z))~~~~~              &  {\rm at~NNLO~(k=2).}
\end{array}
\right.
\end{equation}
This can lead to big positive or negative weights in Monte Carlo 
computations.

\subsection {Behaviour at $z\rightarrow 0$}

The splitting functions
$\{P^V_{\pm},P^S_{\,-}\}$
are logarithmically divergent at $z=0$ starting from the NLO
\begin{equation}
P(\alpha_s,z)\,=\,\sum_{k=0}\alpha_s^{k+1}\,
\bigg\{\sum_{i=1}^{2k}\,{\overline D}_i^{(k)}\,\ln^{i}z\,+\,{\cal O}(1)\bigg\}.
\end{equation}

The remaining splitting functions
$\{P^S_{\,+},P_{FG},\,P_{GF},\,P_{GG}\} $
have the following behaviour:
\begin{equation}
P(\alpha_s,z)\,=\,E_1(\alpha_s)\,\frac{\ln z}{z}\,+\,E_2(\alpha_s)\,
\frac{1}{z}\,+\,
{\cal{O}}(\ln^{2k}\!z )\,,
\end{equation}
The logarithmic term is present starting from the NLO $(k=1)$ 
approximation:
\begin{equation}
E_1(\alpha_s)\,=\,\alpha_s^2\,E_1^{(1)}+\,\alpha_s^3\,E_1^{(2)}\,+\,...\,,
\end{equation}
while the $1/z$ term is present from the LO $(k=0)$ 
approximation
\begin{equation}
E_2(\alpha_s)\,=\,\alpha_s\,E_2^{(0)}+\,\alpha_s^2\,E_2^{(1)}+\,
\alpha_s^3\,E_2^{(2)}...\,.
\end{equation}

\section{Markovian MC for parton-momentum distributions}
\label{sec-markowian-x}

In Ref.~\cite{Golec-Biernat:2006xw} we have described a Markovian MC 
algorithm for {\em parton 
distributions} and we have implemented it in the MC program.
However, the factor $1/z$ in the brems\-strahlung kernels causes
a significant loss of MC efficiency!
We can get rid of this annoying phenomenon by switching
to the $xD(x)$ which evolve with the kernels $zP(z)$.
The reason for improvement is that the kernels $zP(z)$
fulfill  the {\em momentum sum rule}.

The evolution equations for $xD(x)$ read
\begin{equation}
  \label{eq:Evolu1}
  \partial_t\, xD_K(t,x)
   = \sum_J \int\limits_x^1 \frac{d z}{z}\; z\Peu_{KJ}(t,z)\; 
    \frac{x}{z} D_J\Big(t,\frac{x}{z} \Big)\,.
\end{equation}
The kernels 
$\Peu_{KJ}(t,z)\,=\,2\/P_{KJ}(\alpha_s(t),z)$ are split into
virtual and real contributions:
\begin{equation}
  \begin{split}
  \Peu_{KJ}(t,z)&=-\Peu^{\delta}_{KK}(t,\epsilon(t))\, \delta_{KJ}\,\delta(1-z)
                 +\Peu^{\Theta}_{KJ}(t,z),
\\
\Peu^{\Theta}_{KJ}(t,z)&=\Peu_{KJ}(t,z)\,\Theta(1-z-\epsilon(t))\,
\Theta(z-\epsilon'),
  \end{split}
\end{equation}
where $\epsilon$ is an infra-red (IR) cut-off.

The iterative solution obtained from the above formulae reads
\begin{equation}
  \label{eq:Iter6}
  \begin{split}
  xD_K(t,x) =&\, e^{-\Phi_K(t,t_0)} xD_K(t_0,x)
\\&
  +\sum_{n=1}^\infty \;
  \int\limits_0^1 dx_0\;
   \sum_{K_0,\ldots,K_{n-1}}
      \prod_{i=1}^n \bigg[ \int\limits_{t_0}^t dt_i\; \Theta(t_i-t_{i-1})
      \int\limits_0^1 dz_i\bigg]
\\&~~~\times
      e^{-\Phi_K(t,t_n)}
      \prod_{i=1}^n 
          \bigg[ 
                 z_i\Peu_{K_i K_{i-1}}^\Theta (t_i,z_i) 
                 e^{-\Phi_{K_{i-1}}(t_i,t_{i-1})} \bigg]
\\&~~~\times
      x_0 D_{K_0}(t_0,x_0)\, \delta\big(x- x_0\prod_{i=1}^n z_i \big),
  \end{split}
\end{equation}
where $K\equiv K_n$.

The running $\alpha_s(t)$ can be absorbed into the evolution 
variable by the transformation
\begin{equation}
\label{eq:tau1}
t \longrightarrow 
\tau \equiv \frac{1}{\alpha_s(t_A)} \int\limits_{t_A}^{t} dt'\; \alpha_s(t'),
\quad
\frac{\partial t}{\partial\tau}= \frac{\alpha_s(t_A)}{\alpha_s(t)}\,.
\end{equation}
With the choice of $\alpha_s^{(0)}(t)$ in the
definition of $\tau$ and $t_A=t_0$
we get the iterative solution
\begin{equation}
  \label{eq:Iter7}
  \begin{split}
  xD_K(\tau,x) =&\, e^{-\Phi_K(\tau,\tau_0)} xD_K(\tau_0,x)
\\
 & +\sum_{n=1}^\infty \,
  \int\limits_0^1 dx_0
   \sum_{K_0,\ldots,K_{n-1}}
   \prod_{i=1}^n \bigg[ \int\limits_{\tau_0}^\tau d\tau_i\, 
     \Theta(\tau_i-\tau_{i-1}) \int\limits_0^1 dz_i\bigg]
\\&~~\times
      e^{-\Phi_K(\tau,\tau_n)}
      \prod_{i=1}^n 
          \bigg[ \Pcal_{K_i K_{i-1}}^\Theta (\tau_i,z_i) 
                 e^{-\Phi_{K_{i-1}}(\tau_i,\tau_{i-1})} \bigg]
\\&~~\times
      x_0D_{K_0}(\tau_0,x_0)\, \delta\big(x- x_0\prod_{i=1}^n z_i \big),
  \end{split}
\end{equation}
where
\begin{equation}
  \Pcal_{K_i K_{i-1}}^\Theta (\tau_i,z_i)=
   \frac{\alpha_s^{(0)}(t_0)}{\alpha_s^{(0)}(t_i)}\;
    z_i\Peu_{K_i K_{i-1}}^\Theta (\tau_i,z_i)\,.
\end{equation}

In order to generate the above distribution with the MC methods
we simplify the QCD kernels
\begin{equation}
\label{barpcal}
\begin{split}
 &\Pcal^\Theta_{IK}(\tau,z)
  \to \bar{\Pcal}^\Theta_{IK}(\tau_0,z)
   = \Theta(1-z-\bar\epsilon)
       \frac{\alpha_s^{(0)}(t_0)}{\pi} z P^{(0)}_{IK}(z)\,,\\ 
 & z P^{(0)}_{IK}(z)=\frac{1}{(1-z)_+} \delta_{IK} A^{(0)}_{KK}+
            \delta(1-z)    \delta_{IK} B^{(0)}_{KK}
	                              +F^{(0)}_{IK}(z)\,.
\end{split}
\end{equation}
The approximate kernels do not depend on $\tau$!
The compensating weight is
\begin{equation}
  \label{eq:wtZ}
  \bar w_P= \prod_{i=1}^n \frac{  \Pcal^\Theta_{K_i K_{i-1}}(\tau_i,z_i)}%
                           {  \bar\Pcal^\Theta_{K_i K_{i-1}}(\tau_0,z_i)}\,.
\end{equation}

The probability of the forward Markovian leap is now
\begin{equation}
  \label{eq:leap}
  \begin{split}  
    &\bar\omega(\tau_i,x_i,K_i| \tau_{i-1},x_{i-1},K_{i-1})
\\ &
\hspace{20mm}
    \equiv  \Theta(\tau_i-\tau_{i-1})\;
    \bar\Pcal_{K_i K_{i-1}}^\Theta (\tau_0,x_i/x_{i-1})\;
    e^{-\bar{T}_{K_{i-1}}(\tau_i,\tau_{i-1})}\,,
\\ 
    &\int\limits_{\tau_{i-1}}^\infty d\tau_i\; \int\limits_0^1 dz_i \sum_{K_i}
    \bar\omega(\tau_i,x_i,K_i| \tau_{i-1},x_{i-1},K_{i-1})\equiv 1\,,
    \;\; z_i=\frac{x_i}{x_{i-1}}\,.
  \end{split}
\end{equation}

The real-emission form factor is defined as follows
\begin{equation}
\begin{split}
 \bar{T}_K(\tau_{i},\tau_{i-1})
  &= \int\limits_{\tau_{i-1}}^{\tau_{i}} d\tau'\; 
             \int\limits_{0}^{1} dz\;
             \sum_J \bar\Pcal^\Theta_{JK}(\tau_0,z)
\\
  &=  (\tau_{i}-\tau_{i-1})
     \frac{\alpha_s^{(0)}(t_0)}{\pi}
     \bigg[
                     A^{(0)}_{KK} \ln\frac{1}{\bar\epsilon}
      +\sum_{J} \int\limits_0^{1} F^{(0)}_{JK}(z) dz
     \bigg]
\\
  &= (\tau_{i}-\tau_{i-1}) \sum_{J} \bar \pi_{JK} 
   = (\tau_{i}-\tau_{i-1})\; \bar R_K\,.
\\
\end{split}
\end{equation}
On the other hand, the exact virtual (Sudakov) form factor is
\begin{equation}
\label{eq:PhiFF}
  \Phi_K(\tau,\tau_0)
   =\int\limits_{\tau_0}^{\tau} d\tau'\;
   \frac{\alpha_s^{(0)}(t_0)}{\alpha_s^{(0)}(t')}\;
     2
     \left[ A_{KK}(\tau') \ln\frac{1}{\epsilon(\tau')} -B_{KK}(\tau')\right]\,.
\end{equation}

At the LO, for the one-loop $\alpha_s^{(0)}$ and 
$\epsilon(\tau)=\epsilon=const$, it  becomes simply
\begin{equation}
  \Phi_K(\tau,\tau_0)=
    (\tau-\tau_0)\,
     \frac{\alpha_s^{(0)}(t_0)}{\pi}\,
     \left( A_{KK}^{(0)} \ln\frac{1}{\epsilon} -B_{KK}^{(0)}\right)\,.
\end{equation}
At the NLO it is much more complicated, nevertheless 
it can also be integrated analytically, 
see Ref.~\cite{Golec-Biernat:2006xw}.

To complete the Markovianization,
the integral over the ``spill-over'' variable $\tau_{n+1}$
is added
 with the help of the identity
\begin{equation}
\begin{split}
e^{-\Phi_{K_{n}}(\tau,\tau_{n})}
= & \,
  e^{\bar\Delta_{K_{n}}(\tau,\tau_{n})}
\\ 
\times
\int\limits_{\tau}^\infty & d\tau_{n+1}\; 
\int\limits_0^1 dz_{n+1} \sum_{K_{n+1}}
  \bar\omega( \tau_{n+1},x_{n+1},K_{n+1}| \tau_n,x_n,K_n)\,,
\end{split}
\end{equation}
where $z_{n+1}=x_{n+1}/x_n$, and
\begin{equation}
\begin{split}
  \bar\Delta_{K}(\tau_{i},\tau_{i-1}) 
&       = \bar{T}_{K}(\tau_{i},\tau_{i-1})-\Phi_{K}(\tau_{i},\tau_{i-1})
\\
&       = (\tau_{i}-\tau_{i-1}) \bar R_{K}  -\Phi_{K}(\tau_{i},\tau_{i-1})\,.
\end{split}
\end{equation} 
The advantage this method is that
at the LO for $\epsilon=\bar\epsilon$  we obtain
\begin{equation}
  \bar\Delta_{K}=0\,,
\end{equation} 
due to the fact that the kernels obey the {\em momentum sum rule}.
This is also valid at the NLO in the $\overline{MS}$ scheme.
In the actual MC calculations, $\bar\Delta_{K}$ 
can be non-zero due to simplifications in the QCD kernels
at the low MC generation level.

The final formula for this MC scenario with the importance sampling for the
running $\alpha_s$ reads
\begin{equation}
  \label{eq:Markovian4}
  \begin{split}
  x&D_K(\tau,x) 
\\
&
 = e^{\bar\Delta_{K}(\tau,\tau_{0})}
  \int\limits_{\tau_{1}>\tau} d\tau_{1} dz_{1} \sum_{K_{1}}
  \bar\omega( \tau_{1},z_1x,K_1| \tau_0,x,K)\,
  xD_K(\tau_0,x)
\\&
  +\sum_{n=1}^\infty \,
  \int\limits_0^1 dx_0\;
  \int\limits_{\tau_{n+1}>\tau} d\tau_{n+1} dz_{n+1} \sum_{K_{n+1}}\,
  \sum_{K_0,\ldots,K_{n-1}}\;\;
      \prod_{i=1}^n \int\limits_{\tau_i<\tau}^t d\tau_idz_i \,
\\&\times
   \bar\omega( \tau_{n+1},x_{n+1},K_{n+1}| \tau_n,x_n,K_n)\,
   \prod_{i=1}^n 
      \bar\omega(\tau_i,x_i,K_i| \tau_{i-1},x_{i-1},K_{i-1})
\\&\times
      \delta\big(x- x_0\prod_{i=1}^n z_i \big)\,
      x_0 D_{K_0}(\tau_0,x_0)\; \bar w_P\, \bar w_\Delta\,.
  \end{split}
\end{equation}
where 
$z_{i}=x_{i}/x_{i-1},\; K\equiv K_{n}$
and
\begin{equation}
 \label{eq:wt_delta4}
  \bar w_\Delta=e^{\bar\Delta_{K_{n}}(\tau,\tau_{n})} 
      \prod_{i=1}^n e^{\bar\Delta_{K_{i-1}}(\tau_i,\tau_{i-1})}\,.
\end{equation}
For explicit expressions of all ingredients of the above formulae
and for more details see Ref.~\cite{Golec-Biernat:2006xw}.

\section{Numerical tests}
\label{sec-numerics}

We have implemented the above Markovian MC algorithm up to NLO in the
MC program {\tt EvolFMC}.
Then we have performed comparisons of the MC solution of the DGLAP 
with the solutions provided by the non-MC programs  
{\tt QCDnum16}~\cite{qcdnum16} and {\tt APCheb33}~\cite{APCheb33}.
We have evolved the singlet PDF for gluons
and three doublets of massless quarks from $Q_0=1\,$GeV
to $Q=10,\,100,\,1000\,$GeV.
In our test we have used the following parameterization of the starting 
parton distributions in the proton at $Q_0=1\,$GeV:
\begin{equation}
  \begin{split}
    xD_G(x)        &= 1.9083594473\cdot x^{-0.2}(1-x)^{5.0},\\
    xD_q(x)        &= 0.5\cdot xD_{\rm sea}(x) +xD_{2u}(x),\\
    xD_{\bar q}(x) &= 0.5\cdot xD_{\rm sea}(x) +xD_{d}(x),\\
    xD_{\rm sea}(x)&= 0.6733449216\cdot x^{-0.2}(1-x)^{7.0},\\
    xD_{2u}(x)     &= 2.1875000000\cdot x^{ 0.5}(1-x)^{3.0},\\    
    xD_{d}(x)      &= 1.2304687500\cdot x^{ 0.5}(1-x)^{4.0}.
  \end{split}
\end{equation}

\begin{figure}[!ht]
  \centering
  {\epsfig{file=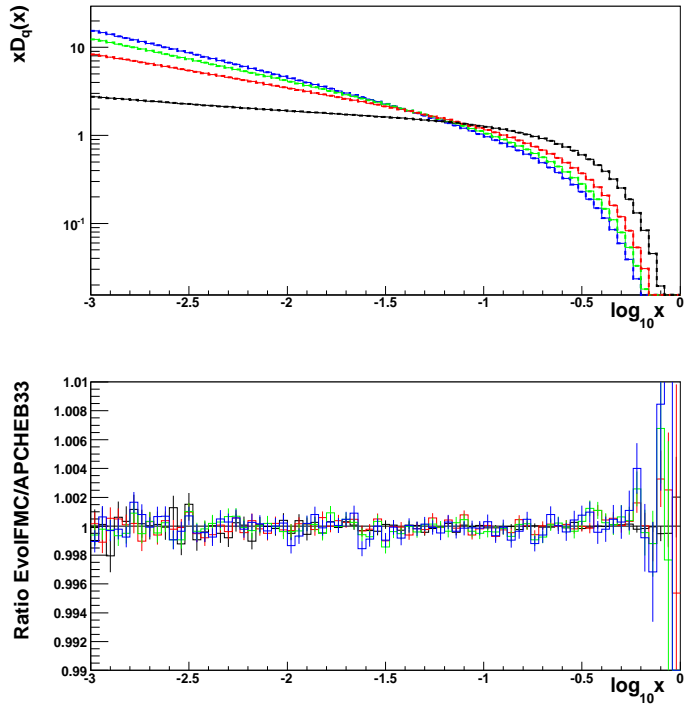,width=100mm,height=100mm}}
  \caption{\sf
    The upper plot shows the quark distribution 
    $xD_q(x,Q_i)$ 
    evolved from $Q_0=1\,$GeV (black)
    to $Q_i=10$ (red), 
    $100$ (green) and $1000$ (blue) GeV,
    obtained in the NLO approximation from {\tt EvolFMC} 
    (solid lines) and  
    {\tt APCheb33} (dashed lines), 
    while the lower plot shows their ratio.
    }
  \label{fig:Gs_NLO}
\end{figure}

\begin{figure}[!ht]
  \centering
  {\epsfig{file=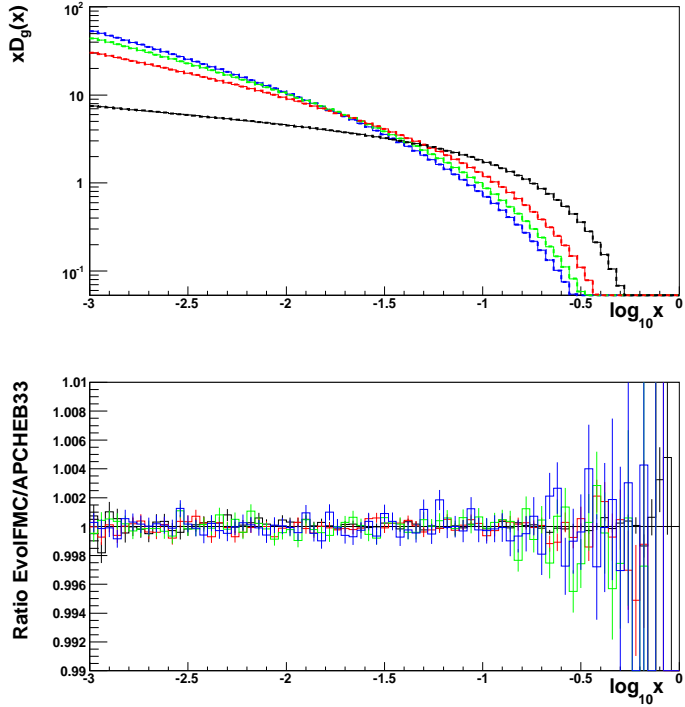,width=100mm,height=100mm}}
  \caption{\sf
    The upper plot shows the gluon distribution 
    $xD_G(x,Q_i)$ 
    evolved from $Q_0=1\,$GeV (black) 
    to $Q_i=10$ (red), 
    $100$ (green) and $1000$ (blue) GeV,
    obtained in the NLO approximation from {\tt EvolFMC} 
    (solid lines) and  
    {\tt APCheb33} (dashed lines), 
    while the lower plot shows their ratio.
    }
 \label{fig:Qs_NLO}
\end{figure}

In Ref.~\cite{Golec-Biernat:2006xw} we have presented the results of
the comparisons between {\tt EvolFMC} and {\tt QCDnum16}
for the gluon and quark-singlet distributions. 
The agreement at the level of $\sim 0.1\%$ has been found for both
the LO and NLO evolution equations.
Here, in Figs.~\ref{fig:Gs_NLO}~and~\ref{fig:Qs_NLO} we show the results of
the comparisons between {\tt EvolFMC} and {\tt APCheb33} for the NLO evolution.
{\tt APCheb33} solves the evolution equations with the use of Chebyshev
polynomials~\cite{APCheb33}. 
As one can see, the gluon and quark-singlet distributions from the two
programs agree within $\sim 0.1\%$
(the similar agreement has been found also at the LO).

\section{Summary and outlook}
\label{sec-summary}

We have constructed the Markovian Monte Carlo algorithm for solving 
the QCD DGLAP evolution equations at the NLO. 
We have implemented this algorithm in the MC program {\tt EvolFMC}
(in C$++$).
We have cross-checked {\tt EvolFMC} with the non-MC programs 
{\tt QCDnum16} and {\tt APCheb33},
and found the agreement at the per-mill level.
MC computation for the NLO evolution is $\sim 5$ times slower
than for the LO evolution.
Singular behaviour of the NLO $P_{FG}$ and $P_{GF}$
splitting functions at
$z\rightarrow 1$ leads to large positive weights for the
$F\rightarrow G$ transitions and to negative weights for the
$G\rightarrow F$ transitions in the region of 
$z\gtrsim 0.95$. This shows the need for additional resummation
in this region.
So far only massless quarks have been considered, however, adding
heavy quarks can be accomplished rather easily.
Also extension to the NNLO seems to be straightforward.
This program can be used as a testing tool for 
constrained MC algorithms for the 
ISR, see e.g.\ 
Refs.~\cite{raport04-06,Jadach:2005bf,Jadach:2005yq,Jadach:2007singleCMC}.
Last but not the least,
this algorithm can form a basis for the FSR parton shower MC event generator. 



\end{document}